# Physics-Aware Style Transfer for Adaptive Holographic Reconstruction

*Chanseok Lee[1], Fakhriyya Mammadova[2], Jiseong Barg[1], and Mooseok Jang[1,3\*]*

[1] Department of Bio and Brain Engineering, KAIST, 291 Daehak-ro, Yuseong-gu, Daejeon, 34141, South Korea.
[2] School of Electrical Engineering, KAIST, 291 Daehak-ro, Yuseong-gu, Daejeon, 34141, South Korea.
[3] KAIST Institute for Health Science and Technology, KAIST, 291 Daehak-ro, Yuseong-gu, Daejeon, 34141, South Korea.

*Correspondence to: mooseok@kaist.ac.kr (M.J.)

**Keywords:** holographic imaging, style transfer, phase retrieval, deep learning

## Abstract

Inline holographic imaging presents an ill-posed inverse problem of reconstructing objects' complex amplitude from recorded diffraction patterns. Although recent deep learning approaches have shown promise over classical phase retrieval algorithms, they often require high-quality ground truth datasets of complex amplitude maps to achieve a statistical inverse mapping operation between the two domains. Here, we present a physics-aware style transfer approach that interprets the object-to-sensor distance as an implicit style within diffraction patterns. Using the style domain as the intermediate domain to construct cyclic image translation, we show that the inverse mapping operation can be learned in an adaptive manner only with datasets composed of intensity measurements. We further demonstrate its biomedical applicability by reconstructing the morphology of dynamically flowing red blood cells, highlighting its potential for real-time, label-free imaging. As a framework that leverages physical cues inherently embedded in measurements, the presented method offers a practical learning strategy for imaging applications where ground truth is difficult or impossible to obtain.

# Introduction

Digital inline holography is a computational imaging technique that reconstructs the complex amplitude of an object from a 2D intensity measurement of a diffraction pattern generated by coherent illumination [1, 2]. Unlike conventional optical microscopy, which relies on absorption or fluorescence contrast, holographic imaging provides label-free, quantitative phase information, enabling high-resolution analysis of morphological and structural features in biological specimens [3-8]. Recent advances in computational algorithms and hardware simplification—such as on-chip, lensless imaging configurations (Fig. 1a)—have enhanced the accessibility and practicality of holographic systems, positioning them as powerful tools for biological research and clinical diagnostics [9-14].

The central challenge in holographic imaging lies in solving an ill-posed inverse problem: recovering the complex-valued object field $x \in \mathbb{C}^{H \times W}$ from an intensity-only measurement $y \in \mathbb{R}^{H \times W}$ recorded by an image sensor. Classical phase retrieval methods, including the Gerchberg–Saxton algorithm, Hybrid Input–Output, transport-of-intensity equation, and multi-height or multi-wavelength techniques, have long been employed to address this problem [15-27]. However, these approaches are fundamentally constrained by slow and unstable convergence, high computational cost, and strong sensitivity to noise and initialization.

To overcome these limitations, recent studies have explored deep learning-based inverse solvers that learn the mapping $G_\theta(y) \to x$ as an approximation of the inverse of the forward imaging process $|F(x)|^2 \to y$, governed by diffraction theory. Essentially, training such models involves learning the inverse function in a data-driven manner by statistically capturing the relationship between the measurement and object domains. In supervised learning, a common approach is to minimize an objective function between the network output $G_\theta(y)$ and the corresponding ground truth $x$ using a large, paired dataset $[x, y]$ [12, 28-44]. This enables fast inference via a single feed-forward pass and supports advanced functionalities such as extended depth-of-field imaging [28], low-photon phase retrieval [31], cross-modality [33], and external generalization [39]. This supervised learning paradigm has also been successfully applied to a wide range of imaging modalities, including fluorescence microscopy [37, 42, 44], structured illumination microscopy [38, 41], and tomography [43].

However, a major limitation across these learning approaches is the challenge of acquiring high-quality object-domain data that is statistically and physically aligned with the measurement-domain data. Obtaining ground truth complex amplitude data typically requires elaborate and expensive optical setups, such as interferometric systems or multi-plane acquisitions. Furthermore, ensuring consistent acquisition parameters, such as numerical aperture (NA), magnification, and illumination conditions, between the object and measurement domains is critical for effective model training, yet often infeasible in practical settings. These constraints severely limit scalability, especially in dynamic or real-time imaging scenarios, and are even more prohibitive in non-optical modalities, such as MRI, X-ray, or electron microscopy, where object-domain ground truth data may be fundamentally unattainable.

In this study, we introduce a style transfer-based phase retrieval framework that learns the inverse mapping function solely from measurement-domain data - specifically, intensity-only diffraction patterns readily obtainable with standard optical microscopes. The key insight is that diffraction patterns inherently encode the object-to-sensor distance through their structural features, such as diffraction rings, which we interpret as a latent physical "style." By leveraging this implicit style information, the proposed adaptive model simultaneously estimates the object-to-sensor distance and reconstructs the complex field of the object, without requiring object-domain ground truth data. Notably, we employ a widely used encoder architecture, VGG-19 pretrained on ImageNet, to extract physical style features, demonstrating that generic computer vision models can be repurposed to capture underlying physical cues embedded in diffraction measurements. We validate the robust performance of our method using both static (3 μm polystyrene beads) and dynamic (flowing red blood cells) experimental datasets. We envision that this approach provides an effective solution for learning inverse mappings in scenarios where object-domain data is inaccessible or costly, with potential applications extending beyond optics to imaging modalities such as MRI, X-ray, and CT.

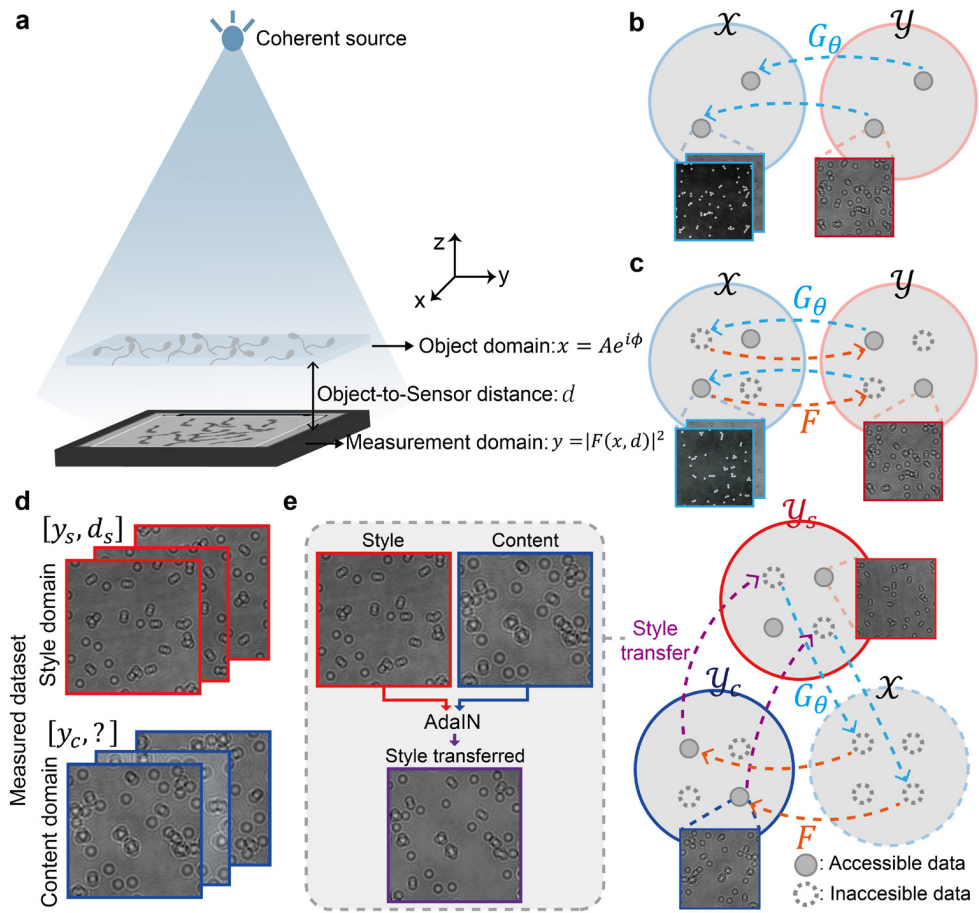

**Fig. 1.** Overall schematic of the proposed method. **a** Digital in-line holographic imaging system. **b,c** Mapping relationship between the object domain $\mathcal{X}$ and the measurement domain $\mathcal{Y}$ in **b** supervised learning and **c** self-supervised learning. **d** The composition of the measurement domain dataset, comprising style domain, $[y_s, d_s]$, which contains intensity measurement and object-to-sensor distance, and the content domain $[y_c, ?]$, which includes only intensity measurement. **e** The mapping relation of the proposed method. The dashed box visualizes the core concept of the proposed physics-aware style transfer framework.

# Principle

## Existing deep learning approaches relying on ground truth data

Fig. 1b and Fig. 1c illustrate two representative learning strategies for solving the inverse problem $x = G_{\theta^*}(y)$: supervised learning with paired datasets and self-supervised learning with unpaired datasets. As shown in Fig. 1b, supervised learning directly compares the network output $G_\theta(y)$ and ground truth $x$ using high-quality, large-scale paired data $[x, y]$, allowing the network to directly capture the statistical relationship between measurements and object-domain representations [12, 28-36, 39, 40]. While effective, this approach requires extensive acquisition efforts and often relies on complex or bulky optical setups, especially when imaging dynamic samples.

Self-supervised methods or neural network-based iterative phase retrieval approaches eliminate the need for paired data by constructing a cyclic mapping between the measurement and object domains [45-51]. As shown in Fig. 1c, the forward model $|F(x,d)|^2 \to y$, governed by known physics, is combined with a learnable inverse model $G_\theta(y) \to x$, forming a cyclic loop that enforces consistency between the object and measurement domains. To address the ill-posedness of the inverse problem and prevent convergence to trivial or non-physical solutions, adversarial regularization is employed by introducing a discriminator to constrain the output distribution of the inverse solver. Minimizing the statistical discrepancy between reconstructed and physically plausible solutions significantly narrows the feasible solution space, enabling stable and effective learning of the inverse operator. While this strategy has shown promising results, it heavily relies on access to ground truth data. For a comprehensive overview of prior work in this area, see Supplementary Note S1.

**Learning inverse mapping without ground truth data**

When access to ground truth object-domain data is limited or fundamentally infeasible, learning an accurate inverse mapping becomes especially challenging, as no explicit information from the object domain can be leveraged for training. To address this limitation, the proposed framework redefines the measurement domain (i.e., intensity domain) by partitioning it into two subsets: a style domain $\mathcal{Y}_s$ and a content domain $\mathcal{Y}_c$. As illustrated in Fig. 1d, the style domain consists of diffraction patterns acquired at known object-to-sensor distances, $[y_s, d_s]$, serving as reference measurements that indicate implicit physical styles embedded in the diffraction pattern associated with specific object-to-sensor distances. In contrast, the content domain contains diffraction patterns for which the object-to-sensor distance is unknown, $[y_c, ?]$. Importantly, the content domain serves as the input domain for the adaptive inverse mapping operation, where a complex-valued map of an object is retrieved from a single content diffraction pattern. It should be noted that the style and content domain images are, in general, unpaired (i.e., not originate from the same objects.)

The core idea of the proposed method is to use the style domain as the intermediate domain to guide the inverse mapping from the content domain to the complex-valued object domain (Fig. 1e). Specifically, the diffraction pattern in the content domain $\mathcal{Y}_c$ is first transformed into an intermediate domain $\mathcal{Y}_s$ (purple arrow in Fig. 1e), which is then mapped to the complex-valued object domain $\mathcal{X}$ through a phase retrieval module (sky-blue arrow in Fig. 1e). Finally, the complex-valued object domain data is mapped back to the content domain $\mathcal{Y}_c$ through a physics-based forward model $F$ grounded in diffraction theory (orange arrow in Fig. 1e) to establish cyclic image translation loop. This approach is motivated by the physical observation that the shape of diffraction patterns such as ring patterns is highly dependent on the object-to-sensor distance. We use adaptive instance normalization (AdaIN) [52] to directly manipulate this physics-grounded variation in the measured patterns as a latent style, and thereby formulating a framework in which style information (i.e. distance information) is transferred from the style domain to the content domain. For details of AdaIN, see "Methods" section.

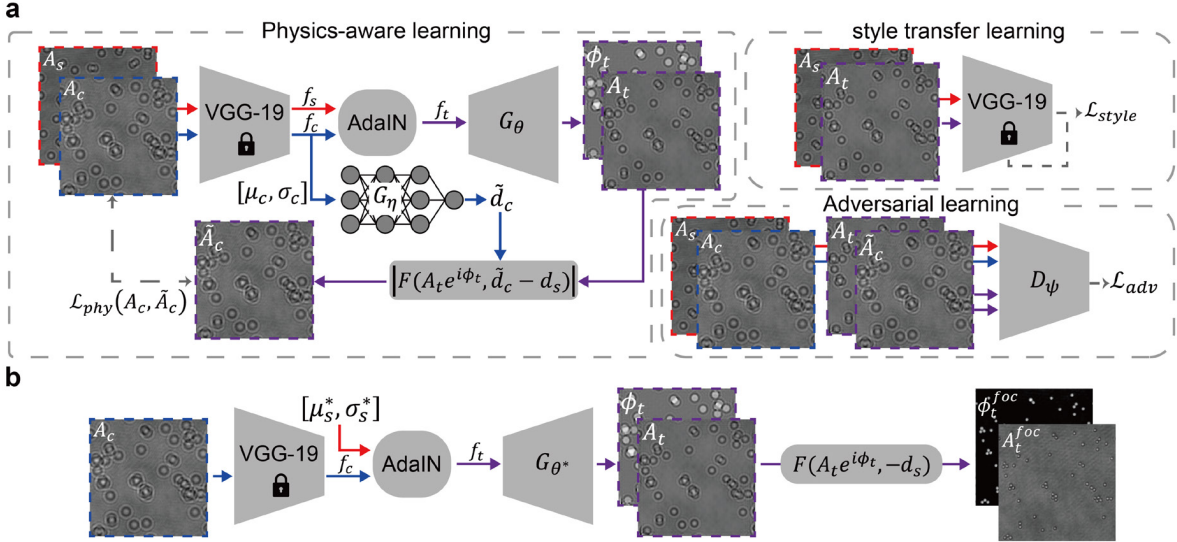

**Fig. 2.** Workflow of the proposed method. **a** Training phase: The network is optimized through a combination of physics-aware, style transfer, and adversarial learning. **b** Inference phase: The reconstructed complex amplitude $A_t e^{i\phi_t}$ is propagated to the focal plane.

Fig. 2a illustrates the training procedure of the proposed style transfer-based phase retrieval framework $S_\theta$, which consists of a VGG-19 encoder $E$ pretrained on ImageNet, followed by AdaIN and a phase-retrieving decoder $G_\theta$. The reconstructed complex field in the object domain $\mathcal{X}_s$ is computed as:

$$A_t e^{i\phi_t} = S_\theta(A_s, A_c) = G_\theta\left(AdaIN(E(A_s), E(A_c))\right). \tag{1}$$

Here, $A_s$ and $A_c$ are the amplitudes of the diffraction pattern from the style and content domains, respectively. By denoting the transferred feature vector $AdaIN(E(A_s), E(A_c))$ as $f_t$, Eq. (1) is simplified to $A_t e^{i\phi_t} = G_\theta(f_t)$.

We employ the style transfer technique with physics-aware learning and adversarial learning in an intertwined manner. The physics-aware learning enforces an identity relation in the cyclic transformation, $\mathcal{Y}_c \to \mathcal{Y}_s \to \mathcal{X}_s \to \mathcal{Y}_c$, where the content diffraction pattern is transferred into an intermediate domain $AdaIN(E(A_s), E(A_c)) \to f_t$. This transferred feature $f_t$ is then passed through the phase retrieval network $G_\theta$ to recover a complex-valued field in the $\mathcal{X}_s$, $G_\theta(f_t) \to A_t e^{i\phi_t}$, which is followed by a physics-based forward model $|F(A_t e^{i\phi_t}, \tilde{d}_c - d_s)| \to \tilde{A}_c$. Notably, leveraging the distance-dependent nature of style information, a shallow network $G_\eta$ is employed to estimate the unknown object-to-sensor distance $\tilde{d}_c$ from the encoded content features. By enforcing the identity relation between the synthesized diffraction pattern $\tilde{A}_c$ and the original input $A_c$, the network is constrained to produce physically consistent reconstructions that satisfy the forward diffraction model. Based on the AdaIN principle, the style transfer aligns the distribution of the amplitude of the retrieved field with that of the style domain by minimizing the discrepancy in the mean and standard deviation between the feature vectors $f_s = E(A_s)$ and $\hat{f}_t = E(A_t) = E(|S_\theta(A_s, A_c)|)$. Finally, adversarial learning regularizes the inverse mapping by ensuring that the amplitude distributions of the transformations $\mathcal{Y}_c \to \mathcal{Y}_s$ and $\mathcal{X}_s \to \mathcal{Y}_c$ resemble those of the style and content domains, respectively. Refer to "Methods" Section and Algorithm 1 for a detailed derivation of the total loss function and a formal description of the proposed method's training procedure.

In the inference stage, the proposed method reconstructs the complex amplitude of the object by propagating the retrieved complex field in $\mathcal{X}_s$ to the focal plane, as illustrated in Fig. 2b. This is expressed as: $A_t^{foc} e^{i\phi_t^{foc}} = F\left(G_\theta\left(AdaIN(f_s^*, E(A_c))\right), -d_s\right)$, where $f_s^* = [\mu_s^*, \sigma_s^*]$ denotes the representative style information computed by averaging the feature statistics $\mu_s^* = E[\mu_s]$ and $\sigma_s^* = E[\sigma_s]$ from a set of style-domain diffraction patterns sampled during training. This representative style vector stabilizes the AdaIN operation and facilitates consistent reconstruction of the complex field at the object-to-sensor distance of $d_s$. When the style domain includes diffraction patterns measured at multiple object-to-sensor distances, separate representative style vectors can be computed for each distance and used to retrieve $A_t^{foc} e^{i\phi_t^{foc}}$. The reconstruction process is summarized in Algorithm 2 in the "Methods" section.

# Results

## Connection between diffraction pattern and style representation

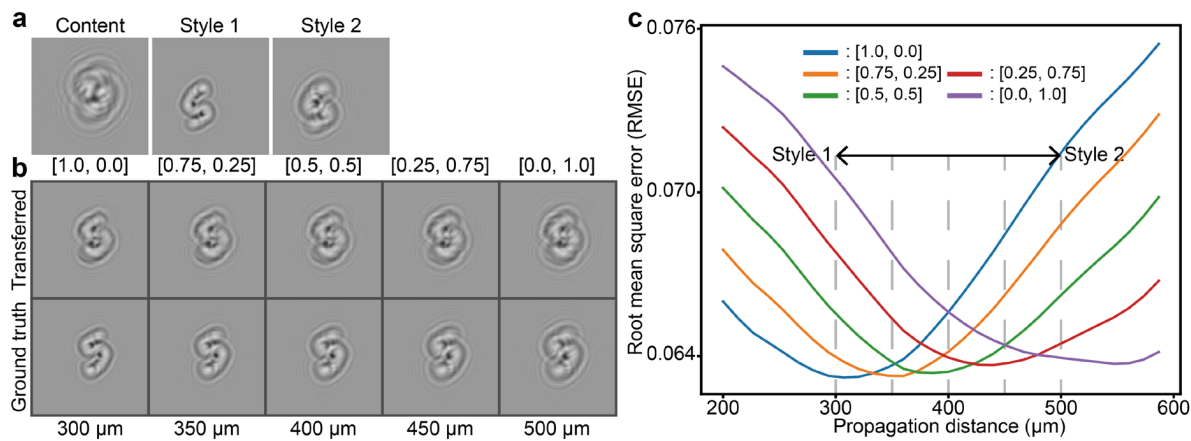

**Fig. 3.** Investigation of the relationship between diffraction pattern shape and style representation. **a** Simulated diffraction patterns from the MNIST dataset, including one content diffraction pattern (800 µm) and two style diffraction patterns acquired at 300 µm (style 1) and 500 µm (style 2). **b** Top: reconstructed diffraction patterns generated by decoding the interpolated latent feature $f_t = \sum_i w_i AdaIN(f_{s_i}, f_c)$ using the trained decoder $G_{\theta^*}$, while varying the weight combination [$w_1$, $w_2$]. Bottom: numerically propagated diffraction pattern. **c** RMSE between each transferred image and the numerically propagated diffraction patterns across a distance range of 200 µm to 600 µm, with a spacing of 20 µm.

To validate the hypothesis that physical distance can be interpreted as an implicit and manipulatable style, we tested the capability of the AdaIN technique of synthesizing the diffraction pattern at intermediate depths. A key strength of the AdaIN technique is its adaptability to embed varying styles into a single content image by a simple operation on the feature vector [52]. Specifically, the feature representation of a content image, $f_c$, can be adaptively transferred with the features from multiple style images, $f_{s_i}$, via the weighted sum $f_t = \sum_i w_i AdaIN(f_{s_i}, f_c)$, where $\sum_i w_i = 1$. This formulation allows the output to reflect a blended style according to the weights $w_i$. If the style representation accurately reflects the underlying distance information, the style information embedded in the transferred feature $f_t$ corresponds to a diffraction pattern at the interpolated distance $d_t = \sum_i w_i d_{s_i}$, enabling the generation of physically meaningful diffraction pattern based on the simple latent space operation.

To demonstrate this capability, we generated diffraction patterns from 50,000 MNIST training images, modeled as phase objects. Data augmentation, including horizontal/vertical flipping and random x–y translations, was applied to increase diversity. Content diffraction patterns were simulated at propagation distances of 600, 700, and 800 μm, while style diffraction patterns were generated at 300, 400, and 500 μm. Examples of a content diffraction pattern at 800 μm and two style diffraction patterns at 300 μm (style 1) and 500 μm (style 2) are shown in Fig. 3a. The decoder $G_\theta$ was trained on this simulated dataset using only the style transfer learning scheme described in "Methods" section.

After training, a single content diffraction pattern was style transferred with two style diffraction patterns captured at different distances of 300 μm (style 1) and 500 μm (style 2). The interpolation weights $[w_1, w_2]$ were varied from $[1.0, 0.0]$ to $[0.0, 1.0]$ in steps of 0.25, gradually decreasing $w_1$ and increasing $w_2$. As illustrated in Fig. 3b, the synthesized diffraction patterns exhibit a smooth transition from the appearance characteristic of style 1 to that of style 2, depending on the relative contributions of each style.

Remarkably, the style-transferred results presented the lowest root-mean-squared error (RMSE) with the diffraction patterns calculated using physics forward model at intermediate distances corresponding to the weight values used for the blended styles, $d_t$ (Fig. 3c). For example, when the weights are set to [0.5,0.5], the resulting style-transferred diffraction pattern closely matches the numerically propagated pattern at 400 μm = 0.5·300 μm + 0.5·500 μm (see green line in Fig. 3c). Importantly, even though the feature representation is encoded using the pretrained VGG-19 that is generally purposed for vision task, the patterns generated from the simple feature space manipulation exhibit physically meaningful characteristics. This strongly confirms that the proposed style transfer framework effectively captures the implicit variation in the measurement domain data in a physics-aware manner, which can be further leveraged as a learnable style representation.

**Reconstruction of complex amplitude and object-to-sensor distance**

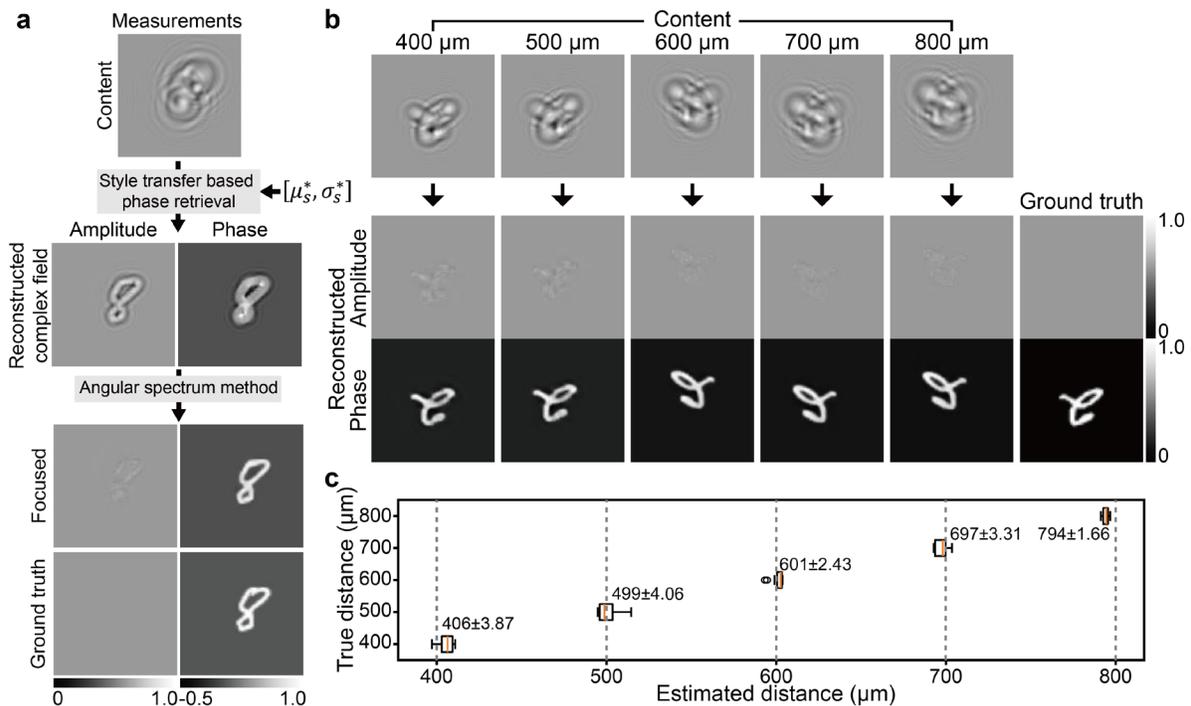

**Fig. 4.** Simulation results of holographic image reconstruction on the MNIST dataset using the proposed method. **a** Overview of the reconstruction pipeline: a content diffraction pattern and a representative style vector are input to the phase retrieval network. The resulting complex field in the object domain $\mathcal{X}_s$ is numerically propagated to the focal plane for final reconstruction. **b** Reconstructed complex amplitudes from content diffraction patterns simulated at propagation distances of 400, 500, 600, 700, and 800 μm. **c** Box plot of object-to-sensor distance estimation for 500 MNIST test samples, with mean and standard deviation shown for each ground truth distance.

To evaluate the effectiveness of the proposed method under conditions where ground truth object-domain data is unavailable, we conducted simulation experiments using the MNIST dataset. A total of 50,000 images were used for training, with data augmentation applied via horizontal/vertical flipping and random x–y translations. The imaging system was modeled with a pixel size of 1.5 μm and a wavelength of 532 nm. Style diffraction patterns were generated at a propagation distance of 200 μm, while content diffraction patterns were generated at 400, 500, 600, 700, and 800 μm. After training, 800 style diffraction patterns were randomly sampled from the training set to compute the representative style feature vector $[\mu_s^*, \sigma_s^*]$. Additionally, 500 content diffraction patterns were generated for testing. Complex amplitude reconstruction was performed following the pipeline described in Algorithm 2 (see Supplementary Note S2.1-2.3 for experimental details, evaluation metrics, and network architectures).

A representative reconstruction result is illustrated in Fig. 4a. The reconstructed complex field in the object domain $\mathcal{X}_s$ was propagated to the focal plane, yielding high-quality focused images that closely matched the ground truth. The reconstructed phase maps achieved a peak signal-to-noise ratio (PSNR) of 29.92 and a mean absolute error (MAE) of 0.0127 rad over 500 MNIST test samples, indicating strong accuracy in recovering both fine structural details and phase information (see Supplementary Fig. S1 for further reconstruction results). Fig. 4b shows reconstructions across varying content distances, highlighting the method's consistent performance across a wide range of propagation depths. Moreover, as shown in Fig. 4c, the distance estimation based on style features achieved an $R^2$ score of 0.9987, validating the effectiveness of the proposed style-based distance inference. Together, these results demonstrate that the proposed method enables accurate phase retrieval and distance estimation solely from diffraction measurements—without requiring ground truth object-domain data.

**Analysis of holographic imaging using 3 μm polystyrene bead**

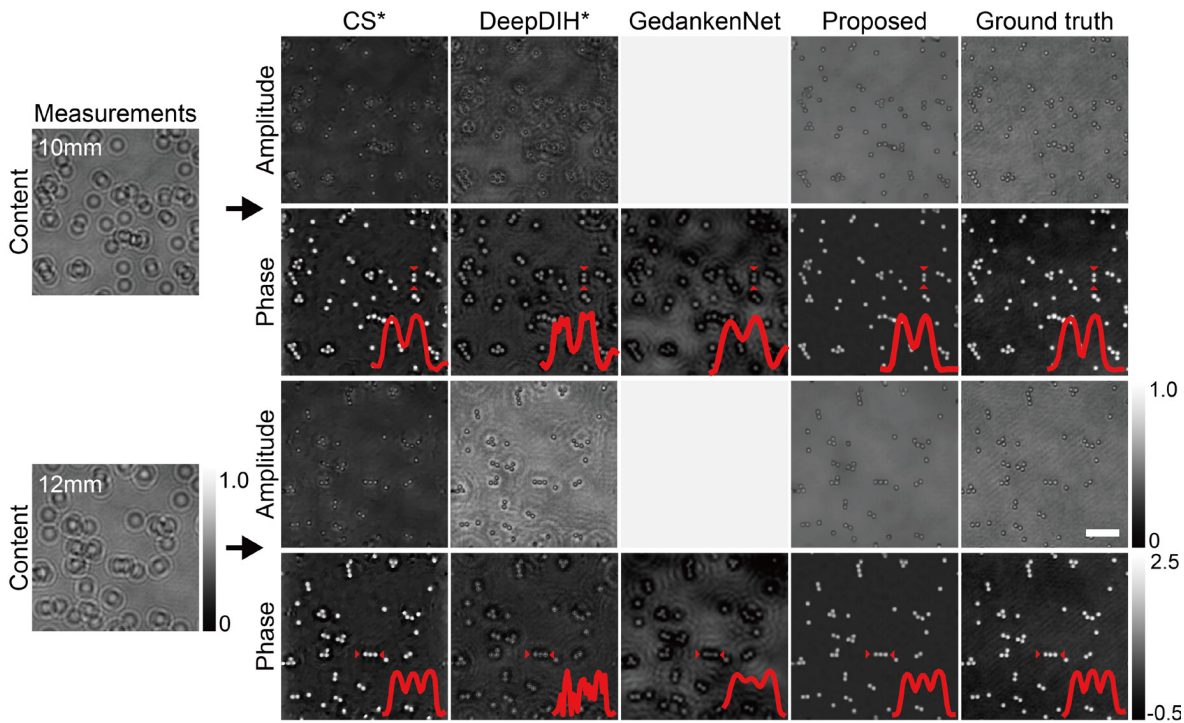

**Fig. 5.** Qualitative comparison of holographic image reconstruction results on the 3 μm polystyrene bead dataset. The representative style vector and content diffraction patterns measured at 10 mm and 12 mm were used as input to the network. Note that amplitude reconstruction is omitted for GedankenNet-Phase, as the method does not estimate amplitude for phase-only objects. Scale bar: 20 μm.

We experimentally validated the reconstruction performance of the proposed method, using a 3 μm polystyrene bead dataset (Polysciences, refractive index 1.60) [50], comparing it against existing methods usable in scenarios where ground truth data is unavailable. In our experiment, the object was relayed through a conventional microscope setup composed of a set of an objective lens and a tube lens, followed by free-space propagation from the image plane to the sensor plane. For training, 4,000 style diffraction patterns captured at 7 mm and 16,000 content diffraction patterns measured at 9, 10, 11, and 12 mm (4,000 per distance) were utilized. Importantly, the distance information of the content domain samples was not used during training. After training, all style diffraction patterns used during training were employed to compute the representative style vector $[\mu_s^*, \sigma_s^*]$. We compared the proposed method against three baseline approaches: classical iterative reconstruction (CS* [17]), untrained neural network-based optimization (DeepDIH* [53]), and a simulation-based self-supervised method (GedankenNet-Phase, N=1 [46]). See Supplementary Note S2.4 for implementation details of the comparison methods.

Fig. 5 presents qualitative comparisons of holographic reconstruction results. It is important to note that CS* and DeepDIH* used object-to-sensor distance as prior knowledge, and GedankenNet-Phase was trained under the expected distance range of 8–13 mm in our experiments. However, despite this prior information, the baseline methods still exhibit noticeable twin-image artifacts and produce blurred reconstructions. In contrast, the proposed method reconstructs complex amplitudes with high spatial fidelity and clear morphological features, closely resembling the expected object structure (see Supplementary Fig. S2 for additional reconstruction results). Quantitative comparisons, summarized in Table 1, show that the proposed method achieves the highest PSNR and the lowest MAE, demonstrating superior reconstruction accuracy. Moreover, the style-based distance estimation achieves an $R^2$ score of 0.9621, validating its robustness in predicting object-to-sensor distance from diffraction patterns alone. The method also benefits from a lightweight architecture and the simplicity of the AdaIN operation, contributing to faster inference times compared to iterative or optimization-based approaches.

**Table 1.** Quantitative comparison of holographic image reconstruction results for 3 $\mu m$ polystyrene bead dataset.

|  | PSNR ↑ | MAE (rad) ↓ | Reconstruction time (s) |
|---|---|---|---|
| **Proposed** | **24.92** | **0.126** | **0.002** |
| CS* | 21.71 | 0.175 | 12 |
| DeepDIH* | 19.64 | 0.221 | 210 |
| GedankenNet-Phase | 17.72 | 0.290 | 0.031 |

Diffraction patterns captured at 9, 10, 11, and 12 mm, with 16 different fields of view at each distance, were used to calculate the evaluation metrics. Bold: best.

**Real-time holographic imaging of red blood cell**

A key advantage of holographic imaging methods is their ability to capture the phase delay introduced by weakly scattering or phase-only objects, such as biological cells, enabling real-time, high-throughput, label-free analysis of dynamic samples [3-5]. This phase delay information is mathematically expressed as $\Delta\phi = 2\pi\Delta n l/\lambda$, where $\Delta n$ is the refractive index difference between the sample and its embedding medium, $l$ is the thickness of the sample, and $\lambda$ is the illumination wavelength, therefore it provides intrinsic contrast that reveals morphological and structural features [4]. Notably, quantitative measurement of $\Delta\phi$ allows for detailed morphological assessment of cells and subcellular structures without the need for staining. For dynamic samples in 3-D environments, it is particularly challenging to capture well-focused complex amplitude maps in the object domain.

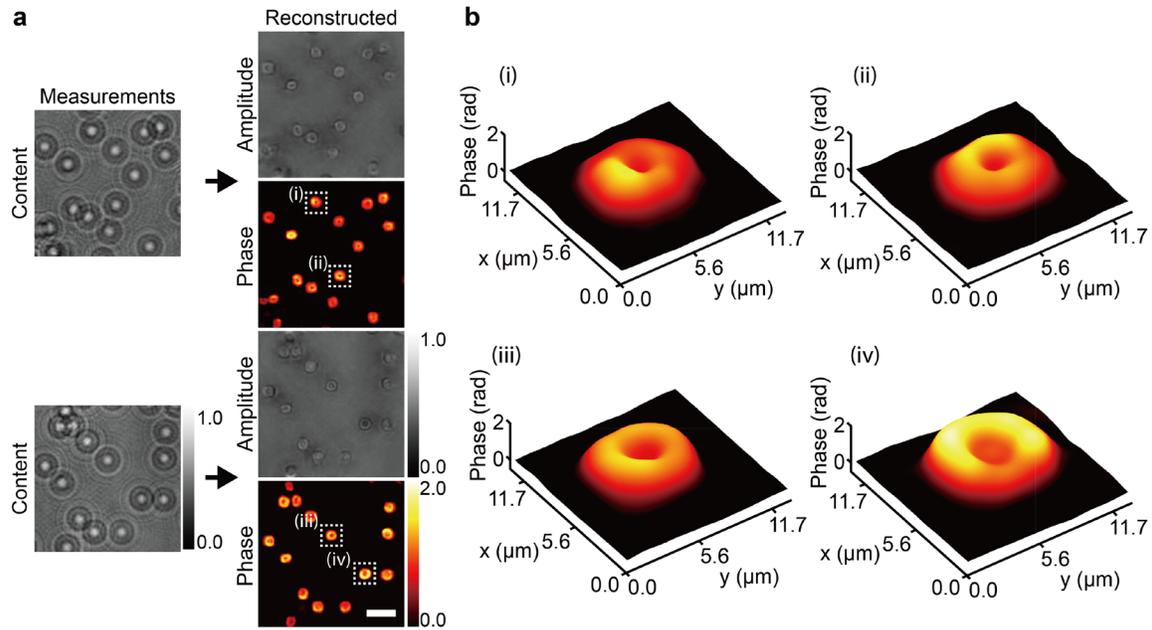

**Fig. 6.** Holographic reconstruction of RBCs flowing in a three-dimensional microfluidic environment. **a** Reconstruction results using the representative style vector and content diffraction patterns with unknown object-to-sensor distances. The reconstructed amplitude and phase images are indicated by arrows. Scale bar: 20 μm. **b** Enlarged 3-D phase maps of selected regions (i–iv), corresponding to the dashed boxes in **a**, illustrating detailed RBC morphology.

We demonstrated the applicability of the proposed method in such dynamic conditions using red blood cells (RBCs) flowing in a 3-D microfluidic environment [50]. The model was trained using 5,000 style diffraction patterns measured at an average object-to-sensor distance of 22.2 mm, along with 6,000 content diffraction patterns with unknown distances. It should be noted that, due to the three-dimensional distribution of the RBCs, not all diffraction patterns in the style domain were captured exactly at 22.2 mm. After training, the representative style vector was computed from all style diffraction patterns used during the training.

The reconstructed complex amplitude maps enabled visualization of both the global flow and the fine morphological features of individual RBCs. As shown in Fig. 6, the proposed method consistently reconstructs the complex amplitudes of RBCs with high fidelity. The estimated phase delay values for the central and peripheral regions of the cells fall within the expected theoretical ranges of 0.47–0.83 and 1.18–2.07 radians, respectively, and the reconstructed three-dimensional morphology closely match these predictions [50]. Supplementary Video 1 further demonstrates successful tracking of RBC dynamics over time. Notably, phase doubling is observed in regions where RBCs overlap, confirming the model's ability to capture additive phase effects and accurately recover physically meaningful phase information. These results highlight the robustness and real-time capability of the proposed framework in dynamic, label-free biological imaging applications.

## Discussion and Conclusion

In this study, we have demonstrated that the style information extracted from the latent space of a pretrained encoder can effectively represent physical parameters implicitly encoded in the diffraction patterns. Specifically, the proposed framework leverages AdaIN to transfer style information from the style domain to the content domain, enabling reliable recovery of the complex object field even in the absence of ground truth data. This approach is grounded in the physical observation that the shape of diffraction patterns is strongly influenced by the object-to-sensor distance, allowing the model to interpret this distance-dependent variation as a latent style representation and use it to guide the learning of the inverse mapping.

Conventional self-supervised approaches to holographic phase retrieval rely on physics-based consistency losses, but without access to object-domain data, they often converge to trivial or degenerate solutions due to the inverse problem's ill-posedness. As illustrated in the results of the ablation study in Supplementary Fig. S3 , a trivial solution such as $x_{triv} = \sqrt{y}e^{i \cdot 0}$ with $d = 0$ may satisfy Eq. (3) but yields no meaningful phase information. To overcome this, we incorporate style transfer and adversarial learning into the phase retrieval framework, using the visual characteristics of diffraction patterns as a proxy for depth. This constrains the solution space and guides the network toward physically valid reconstructions, enabling accurate phase retrieval without requiring object-domain ground truth data.

Importantly, we show that the pretrained VGG-19 encoder—originally trained on natural images from the ImageNet dataset—is capable of extracting meaningful physical cues from diffraction patterns without domain-specific fine-tuning. This indicates that diffraction patterns inherently encode style-like representations, enabling their reuse in physics-informed tasks. As validated in Supplementary Fig. S4, the method performs robust reconstruction across a wide range of object-to-sensor distances, extending beyond the distribution of the training data. Our analysis shown in Supplementary Fig. S5 further reveals that the receptive field of the encoder is critical for interpreting diffraction patterns as style; if the diffraction structure is too large relative to the receptive field, content and style information become entangled within the feature space. Enhancing the encoder's receptive field or adopting localized style-content disentanglement strategies could allow the method to generalize across broader imaging scenarios, including more complex samples and wider propagation distances.

While the current framework is optimized for axially confined samples, it is important to note that diffraction patterns acquired by holographic imaging systems may also encode 3-D structural information. However, extending the proposed method to volumetric samples that span multiple axial planes poses a significant challenge. Specifically, the globally extracted style features from a 2-D diffraction pattern would mix object information from multiple depths, resulting in a single, merged style representation. This blending limits the proposed framework's ability to distinguish and reconstruct depth-specific structures. To overcome this limitation, future work may explore incorporating patch-wise style encoding, multi-scale feature extraction, or advanced disentanglement strategies to enable robust and depth-resolved 3-D phase retrieval.

Overall, we have introduced a novel style transfer-based holographic phase retrieval framework that operates solely on measurement-domain data, eliminating the need for ground truth complex amplitude maps or physical calibration. By removing the dependency on bulky interferometric setups and extensive calibration procedures, this approach substantially reduces the experimental overhead. Notably, the proposed method can be efficiently trained using only the intensity data acquired from standard benchtop microscopes or portable on-chip imaging systems. Furthermore, by bridging the interpretation of physical phenomena with vision-based style transfer techniques, our framework establishes a new paradigm for solving inverse problems in a physics-aware yet data-efficient manner. We anticipate that this approach can be extended to address diverse physical parameters, such as wavelength, polarization, aberrations, coherence, and illumination angle, through vision-guided representations, thereby enabling broader applicability across computational imaging modalities and paving the way toward practical, real-time applications.

## Methods

**Adaptive Instance Normalization (AdaIN) based transformation of diffraction pattern**

Style transfer, a widely adopted technique in computer vision, enables the application of the visual style from a reference image $I_s$ to the content of another image $I_c$, where style typically refers to textures, patterns, or brushstrokes, and content refers to the underlying structural elements or objects [52, 54-56]. In the context of holographic imaging, we draw a direct analogy: the structural motifs observed in diffraction patterns of biological and particulate samples (e.g., red blood cells, sperm, aerosols) can be interpreted as style, while the actual object being imaged constitutes the content. Since the morphology of diffraction patterns is inherently dependent on the object-to-sensor distance, the style information implicitly encodes this physical parameter. Based on this insight, our method leverages style transfer to transform a diffraction pattern acquired in the content domain into a style-aligned representation, mimicking the characteristics of patterns from a known style domain.

To implement style transfer within our framework, we employ Adaptive Instance Normalization (AdaIN) [52]. Feature extraction is performed using a pretrained VGG-19 encoder $E$ from which we obtain the content feature vector $f_c = E(I_c)$ and the style feature vector $f_s = E(I_s)$, corresponding to the content and style diffraction patterns, respectively. The AdaIN operation transfers style by aligning the channel-wise mean and standard deviation of the content feature to those of the style feature:

$$f_t = \mu_s + \sigma_s\big((f_c - \mu_c)/\sigma_c\big) = AdaIN(f_s, f_c) \qquad (2)$$

Here, $\mu_c$ and $\sigma_c$ are the mean and standard deviation of the content feature $f_c$, while $\mu_s$ and $\sigma_s$ are those of the style feature $f_s$. The resulting transformed feature vector $f_t$ is then passed through a pretrained decoder $G_{\theta^*}$ to synthesize the style-transferred image $I_t = G_{\theta^*}(f_t)$. In this context, $I_t$ denotes the transformed diffraction pattern whose structural shape aligns with the style image $I_s$ while preserving the spatial content of the original content image $I_c$.

**Physics-aware style transfer network for phase retrieval**

The training framework of the proposed method consists of three learning schemes: (1) physics-aware learning, (2) style transfer learning, and (3) adversarial learning.

*Physics-aware learning*

Physics-aware learning enforces physical consistency between the reconstructed complex field and the input content diffraction pattern. Specifically, the amplitudes of content diffraction pattern $A_c$ and style diffraction pattern $A_s$ are used as inputs to the encoder $E$, and the transformed feature $f_t$ is generated based on Eq. (2). This transformed feature is then passed through a phase retrieving decoder $G_\theta$, which produces the complex field $A_t e^{i\phi_t}$ where the structural appearance of amplitude aligns with the style domain and the phase is simultaneously retrieved. Since the object-to-sensor distance of the content diffraction pattern is assumed to be unknown, the statistical descriptors $[\mu_c, \sigma_c]$ derived from the content feature are input into a shallow fully connected network $G_\eta$ to estimate the corresponding distance $\tilde{d}_c$. The reconstructed complex field $A_t e^{i\phi_t}$ is then propagated using the physical forward model $F$ with the estimated distance $\tilde{d}_c$ to synthesize the content-domain diffraction pattern $\tilde{A}_c$. The physics-aware loss is computed by comparing the $\tilde{A}_c$ with the original measurement $A_c$, thereby enforcing physical consistency in the learning process.

$$L_{phy}(A_c, \tilde{A}_c) = \|A_c - \tilde{A}_c\|_2^2 = \|A_c - |F(A_t e^{i\phi_t}, \tilde{d}_c - d_s)|\|_2^2 \tag{3}$$

Through this physics-aware learning scheme, the proposed model learns to jointly recover both the phase and the relative object-to-sensor distance of the content diffraction pattern. However, due to the inherent ill-posedness of the inverse problem, a trivial solution such as $A_t = A_c$ and $\tilde{d}_c = 0$, may satisfy the physical consistency constraint defined in Eq. (3), without yielding meaningful reconstruction. To address this ambiguity and better constrain the solution space, we introduce style transfer learning to enforce distance-dependent structural consistency, and adversarial learning to regularize the output distribution and suppress physically invalid solutions.

*Style transfer learning*

Style transfer learning enforces structural alignment between the generated diffraction pattern and the reference style image. Specifically, it ensures that the shape of the generated diffraction pattern resembles that of the style image by comparing the style feature vector $f_s$ with the encoded feature vector of the generated output, i.e., $\hat{f}_t = E(A_t)$. To capture multi-scale style characteristics, the style loss is computed by minimizing the difference between the input style features and generated features extracted from multiple layers of the encoder [52, 55]. The style loss is formulated as follows:

$$L_{style} = \sum_i \|\mu_s^i - \hat{\mu}_t^i\|_2^2 + \|\sigma_s^i - \hat{\sigma}_t^i\|_2^2 \tag{4}$$

, where $\hat{\mu}_t$ and $\hat{\sigma}_t$ are the mean and standard deviation of $\hat{f}_t$, and $i$ indicates the layer of VGG-19 and we use 4 layers - *relu1_1, relu2_1, relu3_1, relu4_1* [52].

*Adversarial learning*

Adversarial learning is integrated to further constrain the feasible solution space of the intermediate-domain amplitude $A_t$ and the physically regenerated content diffraction pattern $\tilde{A}_c$. We employ a Wasserstein Generative Adversarial Network (WGAN), where a discriminator is introduced to distinguish between real and generated distributions, thereby enforcing alignment between the generated outputs and the real measurements. The corresponding WGAN loss [57, 58] is defined as follows:

$$L_{adv} = L_{WGAN} + \lambda_{GP} L_{GP}$$
$$= D_\psi(A_{real}) - D_\psi(A_{fake}) - \lambda_{GP}(\|\nabla_{\hat{A}} D_\psi(\hat{A})\|_2 - 1)^2 \quad (5)$$

, where $D_\psi$ is a discriminator, $A_{real} = [A_s, A_c]$, $A_{fake} = [A_t, \tilde{A}_c]$, $\hat{A} = A_{real} + (1-\alpha)A_{fake}$ and $\alpha \sim [0,1]$, and $\lambda_{GP}$ is a hyperparameter. $L_{WGAN}$ enforces the discriminator $D_\psi$ to learn to differentiate the measurement $A_{real}$ and the generated data $A_{fake}$, while the generator $G_\theta$ aims to produce realistic diffraction pattern by minimizing Eq. (5). $L_{GP}$ ensures the $D_\psi$ satisfy 1-Lipschitz condition, promoting stable training [58].

*Learning framework of the proposed method*

Finally, to further leverage the information of style distance, we apply the total variation (TV) loss $L_{tv}$ to the focused complex field $A_t^{foc} e^{i\phi_t^{foc}} = F(A_t e^{i\phi_t}, -d_s)$. The final loss formulation for the discriminator $L_D$ and the generator $L_G$ are defined as follows:

$$L_D = L_{adv} \tag{6}$$

$$L_G = \lambda_p L_{phy} + \lambda_s L_{style} + L_{adv} + \lambda_{tv} L_{tv} \tag{7}$$

Here, $\lambda_p, \lambda_s$ and $\lambda_{tv}$ are hyperparameters for physics-aware loss, style loss, and TV loss, respectively. To optimize hyperparameters, we manually explored hyperparameter configurations to understand the relationship between different weights and their influence on the results. Based on these insights, we then employed Optuna [59] for automated hyperparameter optimization to maximize performance. The optimization of Eq. (6) and (7) is visually presented in Fig. 2a and formally summarized in Algorithm 1. The official code is available at https://github.com/csleemooo/style_transfer_based_holographic_imaging and see Supplementary Note S2.1 for experimental details

---

**Algorithm 1** Training of style transfer-based phase retrieval network.

---

**Require:** style image $A_s$, style distance $d_s$, content image $A_c$, the number of iterations $N_t$, pretrained VGG-19 $E$, initial network parameters $\theta, \eta, \psi$, and Adam hyperparameters $\beta_1, \beta_2$.

1: **for** $t = 1, \ldots, N_t$ **do**
2:     $f_s = E(A_s), f_c = E(A_c)$
3:     $f_t = AdaIN(f_s, f_c)$
4:     $A_t, \phi_t, \tilde{d}_c = G_{\theta,\eta}(f_t)$
5:     $\tilde{A}_c = |F(A_t e^{i\phi_t}, \tilde{d}_c - d_s)|$
6:     $A_t^{foc} e^{i\phi_t^{foc}} = F(A_t e^{i\phi_t}, -d_s)$
7:     $\hat{f}_t = E(A_t)$
8:     $L_D = L_{disc}$
9:     $\psi \leftarrow Adam(\nabla_\psi L_D, \psi, \beta_1, \beta_2)$
10:    $L_G = \lambda_p L_{phy} + \lambda_s L_{style} + L_{disc} + \lambda_{tv} L_{tv}$
11:    $\theta, \eta \leftarrow Adam(\nabla_{\theta,\eta} L_G, \theta, \eta, \beta_1, \beta_2)$
12: **end for**

*Inference procedure of the proposed method*

---

**Algorithm 2** Inference of style transfer-based phase retrieval network.

---

**Require:** style representation $f_s^* = [\mu_s^*, \sigma_s^*]$, style distance $d_s$, content image $A_c$, pretrained VGG-19 $E$, pretrained phase retrieval network $G_{\theta^*}$

1: $f_c = E(A_c)$
2: $f_t = AdaIN(f_s^*, f_c)$
3: $A_t, \phi_t = G_{\theta^*}(f_t)$
4: $A_t^{foc} e^{i\phi_t^{foc}} = F(A_t e^{i\phi_t}, -d_s)$

---